\documentclass[onecolumn,12pt]{revtex4-2}
\usepackage{amsmath,hyperref,amssymb,graphicx,xcolor}
\usepackage[left=0.6in, right=0.6in, top=0.7in, bottom=0.7in]{geometry}

\usepackage{graphicx}\usepackage{amsmath}
\usepackage{amssymb}
\usepackage{mathrsfs}
\usepackage{dsfont}
\usepackage{float}
\providecommand{\keywords}[1]{\textit{keywords: } #1}
\usepackage{eurosym}

\usepackage{amssymb}
\usepackage{graphicx}
\providecommand{\keywords}[1]{\textbf{\textit{Index terms---}} #1}
\DeclareMathOperator{\Tr}{Tr}
\DeclareMathOperator{\Erf}{Erf}
\usepackage{graphicx}

\begin{document}
\title{\textbf{{\large Decoherence effects in a three-level system under Gaussian process}}}
\author{S. M. Zangi $^{1}$\footnote{A.R and S.M.Z have contributed equally. First authorship is shared between A.R and S.M.Z}, Atta ur Rahman$^{2a}$\footnote{Corresponding author: attapk@outlook.com},  Zhao-xo Ji$^{3}$, Hazrat Ali$^{4}$, Huan-Guo-Zhang$^{3}$}
\address{\small{ $^{1}$School of Physics and Astronomy, Yunnan University, P.O. Box 650500 Kunming, China\\
 $^2$School of Physics, University of Chinese Academy of Science,\\ Yuquan Road 19A, Beijing 100049, China\\
$^{3}$Key Laboratory of Aerospace Information Security and Trusted Computing, Ministry of Education, School of Cyber Science and Engineering, Wuhan University, China\\
$^{4}$Abbottabad University of Science and Technology, P.O. Box 22500 Havellian KP, Pakistan}}
\date{\today}   

\begin{abstract}
When subjected to a classical fluctuating field characterized by a Gaussian process, we examine the {purity} and coherence protection in a three-level quantum system. This symmetry of the three-level system is examined when the local random field is investigated further in the noiseless and noisy regimes. In particular, we consider fractional Gaussian, Gaussian, Ornstein-Uhlenbeck, and power-law noisy regimes. We show that the destructive nature of the Ornstein-Uhlenbeck noise toward the symmetry of the qutrit to preserve {purity and coherence} encoded remains large. Our findings suggest that properly adjusting the noisy parameters to specifically provided values can facilitate optimal extended {purity and coherence} survival. Non-vanishing terms appear in the final density matrix of the single qutrit system, indicating that it is in a strong coherence regime. Because of all Gaussian noises, monotonic decay with no revivals has been observed in the single qutrit system. In terms of coherence and information preservation, we find that the current qutrit system outperforms systems with multiple qubits or qutrits using purity and von Neumann entropy. A comparison of noisy and noiseless situations shows that the fluctuating nature of the local random fields is ultimately lost {when influenced by the classical Gaussian noises}. 
\end{abstract}

\maketitle
\section{Introduction}
In recent decades, there has been considerable progress in quantum information processing and quantum computing \cite{Blais, Paesani, Bennett} {inspired by the designing and enhancement of certain related aspects \cite{X1, X2, X3}.} Quantum coherence has remained one of the most active research areas in quantum information sciences and it has been extensively investigated, yielding significant results and improvements in quantum mechanical protocols \cite{Khalid, Adesso}. Coherence preservation in a quantum system ensures successful transmission and higher efficiency in practical quantum information processing. The concept of super-positioning is central to quantum physics and quantum computing and is referred to as quantum coherence. Coherence {has been also found a requirement} for entanglement and other types of quantum correlations \cite{z1,z2,z3, Hu, Wang, Bloch}.
\par
The entangled and coherent states are not physically separated from their surroundings in a practical sense. Connecting such quantum systems to their surroundings results in a loss of coherence and entanglement due to dephasing effects \cite{Zurek}. This can be caused by a variety of factors in the environment, such as random particle mobility, thermal fluctuations, and various disorders, to name a few. Depending on the defect, the type of system, and the type of system-environment interaction involved, these faulty environments produce a variety of noises, for example, see the {Refs.} \cite{AT1, AT2, AT3, AT4, AT5}. In this context, the local environmental description is preferable because it allows for a more comprehensive investigation of quantum systems with multiple degrees of freedom. Coherence dynamics for decreasing degrading effects have been researched both theoretically and experimentally for a range of quantum systems under varied noisy situations \cite{g1, g2, g3, g4, g5}.
\par
We give a thorough investigation of the coherence preservation for a three-level system under various Gaussian noises. Among the noise types are fractional Gaussian noise ($\mathcal{FG}_n$), Gaussian noise $(\mathcal{G}_n)$, Ornstein Uhlenbeck noise ($\mathcal{OU}_n$), and power-law noise ($\mathcal{PL}_n$). These noises are produced by the particles' usual random motion, which can degrade entanglement and coherence \cite{Koutsoyiannis, Mallick, Benedetti}. The study's primary aim will be to devise effective methods for preventing the deteriorating effects of the corresponding Gaussian noise. In addition, the comparative dynamics of coherence under various noises will be thoroughly investigated.
\par
Non-Markovianity, local environment responses to coherence, entanglement protection, and statistical distinguishability on the density operator space, can be investigated using quantum Fisher information and quantum estimation theory, as discussed in \cite{Yuan, Toth}. The primary goal of the measurements is to determine the quantum Fisher information enhancement of open quantum systems so that unknown environmental parameters can be precisely measured. Important results have been achieved to prevent entanglement and coherence losses using quantum Fisher information and estimation theory for single and three-qubit states in \cite{Javed, KenfackQFI, Lu, Li}. In addition, in \cite{Streltsov}, coherence measurement has been investigated using entanglement-based coherence measures, distance-based coherence, and geometric coherence measures. In this paper, we look at the coherence preservation for a single qutrit state using two metrics: purity and von Neumann entropy \cite{g5}. To summarize the noise-free and noisy local fields comprehensive coherence evaluation, besides purity and von Neumann entropy, we will be using {$\ell_1$-norm} of coherence \cite{L1-NORM}.
\par
The phase of a quantum system is crucial to the dynamics and symmetry of quantum systems and associated transmitting channels. After computing the time-evolved density matrix, an average of the noise phases will be calculated to determine the noise's damaging effects. The quantum system's dynamics will be performed using the time unitary operation. Under the classical fluctuating field, the stochastic Hamiltonian is used to describe the energy state of the qutrit system. In both noisy and noise-free local environments, we examine the qualitative behavior of the system's {coherence}. The existence of a pure noiseless configuration is an ideal example; however, to decode a coherent state, it will be necessary to estimate the dissipation power and other local environmental characteristics. Differentiating between noiseless and noisy classical channels could help with quantum mechanical circuit design and long-term coherence preservation. 

\par
The major benefit of Gaussian processes is that they are tractable in many contexts and mimic a wide range of situations reasonably well. For example, in finance, Kalman filters, econometrics, satellite tracking, neural networks, machine learning, Bayesian reasoning, and other fields, Brownian motion and the related Gaussian noises are commonly used \cite{Nirwan, Lazaro, Schwab, Sharifzadeh, Ibrahim, Rodrigues}. $\mathcal{FG}_n$, a stationary time series model with longer memory properties that have been applied in econometrics, hydrology, climatology, functional MRI, traffic networking and signaling, and so forth, could be part of the Gaussian process \cite{Taqqu, Prasad, Pelletier, Maxim, Paxson}. The Ornstein–Uhlenbeck process is stable, Gaussian, and Markovian, with Brownian motion-associated disorder causing $\mathcal{OU}_n$. This noise has been employed in studies including intraday pairs trading strategy, stimulated Raman adiabatic passage systems, quantum control, modern quantum technologies, and other areas \cite{Moura, Blekos, Stefanatos, Stefanatos1}. Besides, $\mathcal{PL}_n$ is another Gaussian noise used to model signal detection, human observer detection experiments, surface growth and dynamics in lipid bilayers \cite{Burgess, Burgess1, Lam, Molina}. The dephasing effects generated by these Gaussian disturbances on the dynamical map of the three-level system in local external fields will be the focus of this paper. The purpose is to characterize the preservation of coherence and information encoded initially in the system over the complete range of the related noisy parameters. In this domain, workable methods for avoiding or minimizing Gaussian dephasing effects will be presented by utilizing the noise phase via noise parameter adjustments. {Therefore, will be leading to longer preservation of quantum correlations. For example, by utilizing the parameters of bit-phase flip and the phase flip channels, quantum correlations have been shown to remain preserved in a noisy accelerated two-qubit system \cite{R1}.} We also intend to provide computational values for the $\beta$-functions concerning various noise parameter values, which define the superimposed noise phase over the collective phase of the system and a classical fluctuating environment. This will lay the groundwork for practical optimization of the classical fields with relative noise disorder. 

The paper is organized as: In Sect.\ref{sections2}, the physical model and associated dynamics of the three-level system in the classical fluctuating field under distinct noises will be given. In Sect.\ref{section3}, results obtained are discussed. Sect.\ref{section4} will explain the conclusion from the present investigations.

\section{Model and dynamics}\label{sections2}
This section summarizes the required mathematical operations. We focus on the dynamical map of coherence in a three-level system that exists initially and then evolves in classical fields of fluctuating nature driven by distinct Gaussian noises. The Hamiltonian that governs the system's present dynamical map can be stated as \cite{g5}:
\begin{equation}
\mathcal{H}_{qt}(t)=\varepsilon I+\omega \eta(t)S_x, \label{Hamiltonian}
\end{equation}
\begin{figure}[ht]
\centering
\includegraphics[width=7cm, height=7cm]{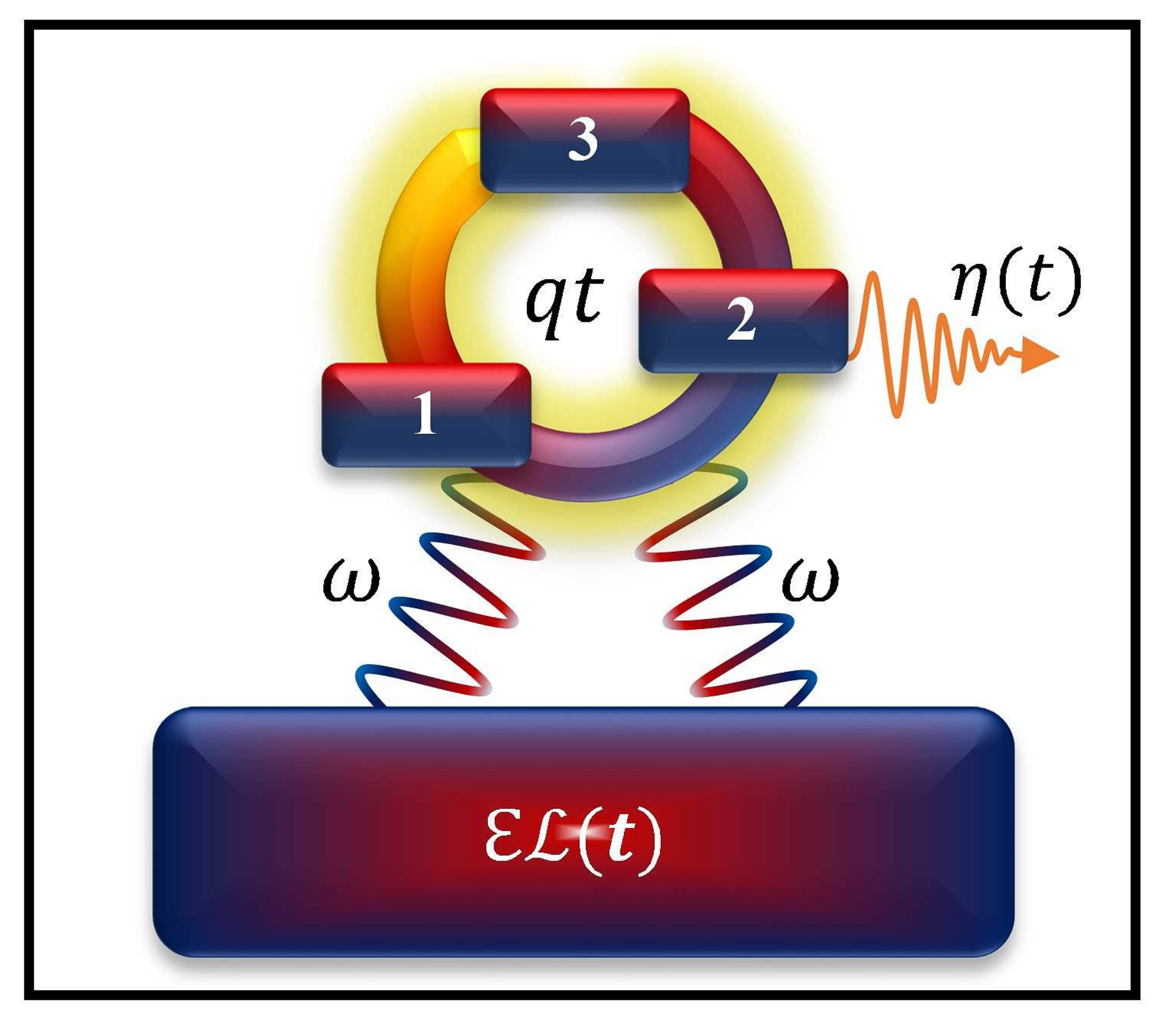}
\caption{The current configuration model depicts the coupling of a three-level system $qt$ exposed to a classical fluctuation field $\mathcal{EL}(t)$. The system-environment coupling strength $\omega$ is shown by the blue-reddish wavy lines, while the noise's influence is represented by the yellowish light in the qutrit. The brownish-wavy lines depict system dynamics as defined by the associated environment's stochastic parameter $\eta(t)$, with diminishing amplitude showing Gaussian noise-induced dephasing.}
\end{figure}
where $\varepsilon$ is the energy associated with the system. The qutrit's space is defined by the identity matrix $I$ and the spin-$1$ operator $S_x$. $\omega$ is the qutrit-environment coupling strength while $\eta(t)$ is the stochastic parameter that flips between $\pm1$ and its nature depends on the classical noise imposed on the external fields. Because of the random nature of the $\eta(t)$, the given Hamiltonian in Eq.\eqref{Hamiltonian} is stochastic and explains the stochastic dynamics of the three-level system with time. The corresponding time evolution operation for the three levels system can be written as \cite{Rossi}:
{\begin{align}
 \mathcal{U}_{qt}(t)&=\exp\left[-i\int_0^t \mathcal{H}_{qt}(s)ds\right],
\end{align}}
where $\hbar=1$. If the qutrit system is initially prepared in the state $\rho_o$, then the time-evolved density matrix can be put into the following form as \cite{Rossi}:
\begin{align}
\rho_{qt}(t)&=\mathcal{U}_{qt}(t) \rho_o {U_{qt}(t)}^\dagger.\label{final density matrix}
\end{align} 
\subsection{Impact of local Gaussian noises}
{Various local Gaussian noises are defined here to include the noisy effects.} In this proceeding, we evaluate the application of the fractional Gaussian ($\mathcal{FG}_n$), Gaussian $(\mathcal{G}_n)$, Ornstein Uhlenbeck $(\mathcal{OU}_n)$ and power-law noise $(\mathcal{PL}_n)$. The random mobility of the particles in the diffusion process causes $\mathcal{FG}_n$ and $\mathcal{OU}_n$. This can disrupt the dynamics of quantum systems, necessitating further investigation. Because of its larger timescale correlations, $\mathcal{FG}_n$ has been utilized to analyze meteorological information \cite{Masoomy}, traffic control analysis \cite{Ledesma}, and electrical measurements \cite{Luft}. The $\mathcal{OU}_n$ has been extensively studied with dynamics of quantum systems, as described in \cite{Rossi}. Similarly, the discrete nature of warm object radiation causes $\mathcal{G}_n$ and gives rise to thermal vibration of the medium's particle and is used to study digital imaging \cite{Saravanan}, signal detection, and phase transitions \cite{Merhav}. Finally, in solid-state and superconducting materials, the $\mathcal{PL}_n$ is a low-frequency noise caused by resistance. This noise has been already studied for scaling of surface fluctuations and dynamics of surface growth models \cite{Amar}, human observed detection experiments with mammograms \cite{Burgessx}, discrete simulation of colored noise and stochastic processes \cite{Kasdin}.

\par To impose classical noise over the time evolution of the system, one has to include $\beta$-function, which reads as \cite{Rossi}:
\begin{equation}
\beta(t)=\int_0^t \int_0^t K(s-s^{\prime})ds ds^\prime. \label{Beta function}
\end{equation}
We look at four different Gaussian processes. To be more specific, we suppose that the stochastic field $\beta(t)$ is driven by $\mathcal{FG}_n$, $\mathcal{OU}_n$, $\mathcal{G}_n$, or $\mathcal{PL}_n$. The associated autocorrelation function {of} $\mathcal{FG}_n$ with a diffusion coefficient which proportionally grows as $\tau^{2H}$ can be written as \cite{Benedetti, Rossi}:
{
\begin{align}
K_{\mathcal{FG}_n}(\tau-\tau^{\prime})=&\frac{|\tau^\prime|^{2H}-|\tau-\tau^\prime|^{2H}+|\tau|^{2H}}{2}.\label{AC of FN}
\end{align}}
where $0 \leq H\leq 1$ is known as the Hurst index ($H$). This autocorrelation function is specifically defined for three different values. When $H=\frac{1}{2}$, the effects of $\mathcal{FG}_n$ noise becomes similar to the Wiener process. At {$H < \frac{1}{2}$} the Eq.\eqref{AC of FN} enters a sub-diffusive process and the increments of the expressions are negatively correlated. When {$H > \frac{1}{2}$}, the autocorrelation function of the noise reaches the super-diffusive regime and the related increments of the equation become positive. The $\beta$-function for the $\mathcal{FG}_n$ can be obtained by putting the auto-correlation function from Eq.\eqref{AC of FN} into Eq.\eqref{Beta function} as \cite{Rossi}:
\begin{equation}
\beta_{\mathcal{FG}_n}(\tau)=\frac{\tau^{2(H+1)}}{2(H+1)}.\label{Beta function of FN}
\end{equation}
In the case of $\mathcal{G}_n$, $\mathcal{OU}_n$ and $\mathcal{PL}_n$, the corresponding autocorrelation expressions are:
\begin{align}
K_{\mathcal{G}_n}(t-t^\prime,& \gamma,\Gamma)=\frac{\Gamma\gamma \exp[-\gamma^2(t-t’)^2]}{\sqrt{\pi}},\label{AC of GN}&  \\
K_{\mathcal{OU}_n}(t-t^{\prime},&\gamma,\Gamma)=\frac{\gamma \Gamma \exp[-\gamma |t-t^\prime|]}{2}, \label{AC of OU}&\\
K_{\mathcal{PL}_n}(t-t^{\prime},&\gamma,\Gamma,\alpha)=\frac{[\alpha-1]\chi \Gamma}{2[\chi \vert t-t^{\prime}\vert+1]^2}, \label{AC of PL}& 
\end{align}
{where $\Gamma$ regulates the damping rate, $\alpha$ is the unknown parameter in the case of $\mathcal{PL}_n$ which has been also demonstrated to carry noisy detrimental effects. For the $\mathcal{G}_n$, we assume $g=\frac{\gamma}{\Gamma}$ and $ \tau=\Gamma t$ where $\gamma$ is also a dephasing Gaussian noise parameter. By putting the autocorrelation function from Eq.\eqref{AC of GN} into Eq.\eqref{Beta function}, we get} \cite{Benedetti}:
\begin{equation}
\beta_{\mathcal{G_N}}(\tau)=\frac{1}{g}\left[\frac{e^{-g^2\tau^2}-1}{\sqrt{\pi}}+\Erf[g\tau](g \tau)\right],\label{Beta function of GN}
\end{equation}
where {$\Erf[g\tau]=\frac{2}{\sqrt{\pi}}\int_0^{g\tau} \exp[-t^2]dt$ is the error function of the normalized Gaussian distribution}. The relative $\beta$-function for the $\mathcal{OU}_n$ can be obtained by inserting Eq.\eqref{AC of OU} into Eq.\eqref{Beta function} and can be put into the form \cite{Rossi}:
\begin{equation}
\beta_{\mathcal{OU}_n}(\tau)=\frac{g\tau+\exp[-g\tau]-1}{g},\label{Beta function of OU}
\end{equation}
where $g$ is the inverse of the autocorrelation time $\tau$. Similarly, for $\mathcal{PL}_n$, we {recall $g=\frac{\gamma}{\Gamma}$ and $ \tau=\Gamma t$ and insert} the autocorrelation function given in Eq.\eqref{AC of PL} into Eq.\eqref{Beta function}, we get the $\beta$-function as \cite{Benedetti}:
\begin{equation}
\beta_{\mathcal{PL}_n}(\tau)=\frac{g\tau (\alpha -2)-1+(1+g\tau )^{2-\alpha }}{g(\alpha -2)}.\label{Beta function of PLn}
\end{equation}
The final density matrix of the system is averaged over the random phase factor $\phi$ as $\langle e^{\pm n \phi(t)}\rangle = \langle e^{\theta(\tau)} \rangle$ where $\phi=i n \omega \eta(t)$ with $n \in \mathbb{N}$ is the phase factor determining the dynamical characteristic of the system in the local random fields and $\theta(\tau)= -\frac{1}{2}n^2 \beta_{\mathtt{\mathcal{AB}}}(\tau)$ {(with $\mathcal{AB} \in \{\mathcal{FG}_n,~ \mathcal{G}_n,~\mathcal{OU}_n,~\mathcal{PL}_n$)} is the phase factor of the local Gaussian noise. The time-evolved density matrix of the system can be obtained by \cite{Rossi}:
\begin{equation}
\rho_{qt}(\tau)=\left\langle U_{qt}(t)\rho_o U^{\dag}_{qt}(t)\right\rangle_{\theta(\tau)}.\label{final density matrix under Gaussian process}
\end{equation}
\subsection{Coherence measures}
The degree of mixedness of a quantum state, or the coherence of a pure initial state that reflects the physical system because of interactions with classical environments, is determined by purity. For a quantum state $\rho_{qt}(\tau)$, purity can be determined by \cite{g5}:
\begin{equation}
\mathcal{P}_r(\tau)=\hbox{Tr}\left[\rho^2_{qt}(\tau)\right], \label{purity}
\end{equation}
where for a system of n-dimensions, purity ranges as: $\frac{1}{n} \leq \mathcal{P}_r(\tau) \leq 1$. The state of being pure and coherent occurs at $\mathcal{P}_r(\tau)=1$, while entirely mixed and decoherent at $\frac{1}{n}$.
\par
Decoherence occurs when quantum systems' wave functions interact with their coupled environments. As a result, rather than being a single coherent quantum superposition, the system behaves like a classical statistical ensemble of its constituents. The decoherence phenomenon will be a credible measure to calculate coherence loss in the time-evolved state of the system because the system-environment interaction is depicted classically here. The von Neumann entropy technique can assess the decoherence effects for the time-evolving density matrix as \cite{g5}:
\begin{equation}
\mathcal{V}_e(\tau)=-\Tr\left[ \rho_{qt}(\tau) \log\rho_{qt}(\tau)\right]. \label{von Neumann entrophy}
\end{equation}
$\mathcal{V}_e(\tau)=0$ indicates the state is coherent with no information loss. Any other value will represent the corresponding amount of coherence and information loss for the three-level system.

\section{Main Results}\label{section3}
In this part, we provide the major findings for the dynamics of a qutrit system under Gaussian noise emanating from a classical fluctuating field. Apart from that, the results of Eqs.\eqref{purity} and \eqref{von Neumann entrophy} will be analyzed to see how purity and von Neumann entropy have changed over time. In order to determine the current noise's ability to dephase the single qutrit system, the system's dynamics are explored in both noiseless and noisy conditions.
\subsection{Noiseless classical field}
The dynamics of the single qutrit system, when subjected to a noise-free stochastic field is discussed in this section. Using the existing dynamical setup of the system in a noise-free context, the original role of stochastic fields in noise exclusion can be demonstrated. This will also discriminate between the qualitative dynamical map of a single qutrit system in both noiseless and noisy classical fields. The following is the time unitary operation matrix:
\begin{align}
\mathcal{U}_{qt}(t)&=\exp(-i t \epsilon)\left[
\begin{array}{ccc}
 \cos\left[\frac{\phi }{2}\right]^2 & -\frac{i  \sin[\phi ]}{\sqrt{2}} & \frac{1}{2}  (-1+\cos[\phi ]) \\[3mm]
 -\frac{i  \sin[\phi ]}{\sqrt{2}} &  \cos[\phi ] & -\frac{i  \sin[\phi ]}{\sqrt{2}} \\[3mm]
 \frac{1}{2}  (-1+\cos[\phi ]) & -\frac{i  \sin[\phi ]}{\sqrt{2}} &  \cos\left[\frac{\phi }{2}\right]^2
\end{array}
\right],
\end{align}
{while the initial density matrix is considered in the form} $\rho_o=\vert \psi \rangle \langle \psi \vert$, $\psi=\frac{1}{\sqrt{3}}(\vert0\rangle+\vert1\rangle+\vert2\rangle)$. The final density matrix, computed using Eq.\eqref{final density matrix}, can be put into the following form:
\begin{align}
\rho_{qt}(t)&=\frac{1}{12}\left[
\begin{array}{ccc}
  3+\cos[2 \phi ] &  4+i \sqrt{2} \sin[2 \phi ] &  3+\cos[2 \phi ] \\[2mm]
 4-i \sqrt{2} \sin[2 \phi ] & 6-2\cos[2 \phi ] & 4-i \sqrt{2} \sin[2 \phi ] \\[2mm]
 3+\cos[2 \phi ] &  4+i \sqrt{2} \sin[2 \phi ] &  3+\cos[2 \phi ]
\end{array}
\right].\label{final density matrix value}
\end{align} 
The structure of the density matrix given in Eq.\eqref{final density matrix value} illustrates that the system is coherent. As the off-diagonal elements of the matrix are non-zero. Besides, for the density matrix $\mathcal{P}_r(t)=1$ and $\mathcal{V}_e(t)=0$ hence, portray the system to be coherent.
\par
To analyze the coherence qualitative dynamical map for the time evolved density matrix of the system given in Eq.\eqref{final density matrix value}, we use the $\ell_1$-norm of coherence, which can be computed as \cite{L1-NORM}:
\begin{equation}
C(t)=\sum_{i \neq j}\vert \langle i \vert \rho_{qt}(t) \vert j \rangle \vert,\label{L1 norm coherence}
\end{equation}
where $\sum_{i \neq j}\vert.\vert$ is the sum of the absolute values of the off-diagonal elements of $\rho_{qt}(t)$ in the chosen reference basis. {It is worth mentioning that the $\ell_1$-norm of coherence has range $0 \leq C(t) \leq n-1 $ where $n$ is the dimensions of the system \cite{d1}. Therefore, for the current qutrit system $\ell_1$-norm of coherence will range as $0 \leq C(t)  \leq 2 $.}
\par
	\begin{figure}[!h]
		\begin{center}
			\includegraphics[width=0.40\textwidth, height=160px]{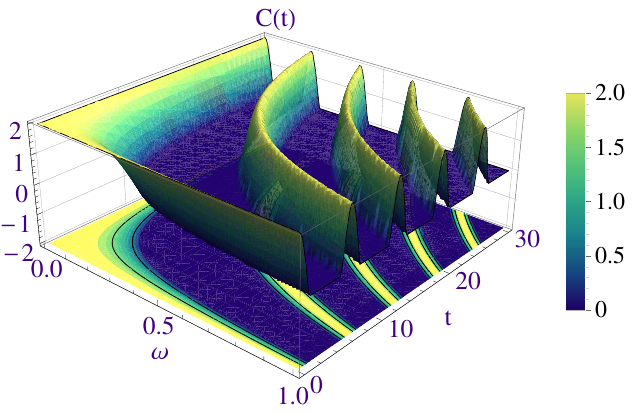}
			\put(-210,160){($ a $)}\quad
			\includegraphics[width=0.40\textwidth, height=160px]{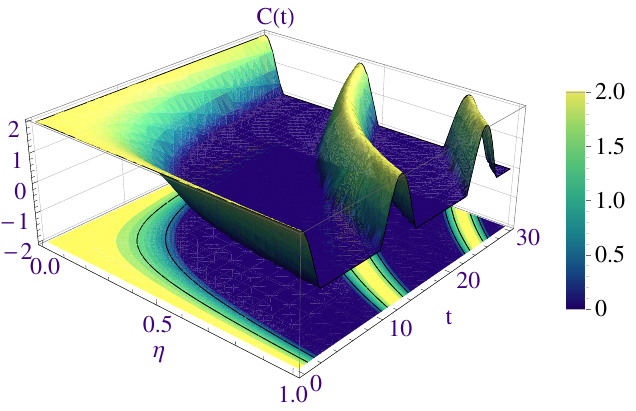}
			\put(-210,160){($ b $)}\quad
		\end{center}
\caption{Time evolution of coherence in a single qutrit system prepared in the time evolved state $\rho_{qt}(t)$ given in Eq.\eqref{Matrix with applied noises} subjected to a noiseless classical channel when (a) $\eta=1$, $0 \leq \omega \leq 1$ and (b) $0 \leq \eta  \leq 1$, $\omega=0.5$ against time evolution parameter $t$.} \label{Fluctuations without noise}
\end{figure} 

$Fig.\ref{Fluctuations without noise}$ shows the dynamics of coherence for the time-evolved density matrix state given in $Eq.\eqref{final density matrix}$ for the single qutrit system when coupled to the classical field {against coupling strength $\omega$ and classical stochastic parameter $\eta$.} Note that the superposition of the noise phase over the system phase is not applied. From the current results, we find that such local environments have a random character and strongly support coherence revivals. These fluctuations are the main reasons for the dynamics and preservation of coherence. As in most of the previous results, whenever the revival character in the dynamical maps of the system vanishes, the coherent states become decoherent, for example, see the Refs. \cite{Mazzola, Rossi, BenedettiC, Kenfack-CN, y1}. Thus, the fluctuating character of the current local channels influenced by disorders can be helpful to preserve coherence encoded initially in a quantum state. Here, the fluctuation rate in coherence is influenced by the {qutrit-environment coupling strength $\omega$ and stochastic parameter $\eta$ of the classical field}. As one can note that the number of revivals is greater for the increasing strength of $\omega$ in $Fig.$\ref{Fluctuations without noise}(a) and for $\eta$ in $Fig.$\ref{Fluctuations without noise}(b). In agreement, the stochastic parameter also influences the revival frequency of coherence, as for the increasing $\eta$ values, the $\ell_1$-norm of coherence faces repeated and increasing number of oscillations. The amplitude of the fluctuations is completely independent of the $\omega$ as well as of $\eta(t)$ and only depends upon the type of system involved. This optimal setting for controlling the revivals of coherence can lead to engineering and designing circuits and protocols for required results \cite{Benedetti-designing circuits, Sweke, Ji}. We conclude that the three-level system remains a resource state in the noise-free classical fields, exhibiting a dynamical map with no coherence loss and that it does not transition from the resource state to the free state regime completely.
\subsection{Classical field with Gaussian noises}
The application of Gaussian noises to the time-evolved density matrix of the single qutrit system described in Eq.\eqref{final density matrix under Gaussian process} is covered in this section. In the current case, the noise phase is superimposed over the system's phase. The final density matrix for the three-level system under the Gaussian noise computed has the form:
\begin{align}
\rho^{\mathcal{AB}_n}_{qt}(\tau)=\left[
\begin{array}{ccc}
\frac{1}{12} \left(3+\mathcal{M}_{\mathcal{AB}_n}\right) & \frac{1}{3} & \frac{1}{12} \left(3+\mathcal{M}_{\mathcal{AB}_n}\right) \\[2mm]
 \frac{1}{3} & \frac{1}{2}-\frac{\mathcal{M}_{\mathcal{AB}_n}}{6}  & \frac{1}{3} \\[2mm]
 \frac{1}{12} \left(3+\mathcal{M}_{\mathcal{AB}_n}\right) & \frac{1}{3} & \frac{1}{12} \left(3+\mathcal{M}_{\mathcal{AB}_n}\right) \\
\end{array}
\right],\label{Matrix with applied noises}
\end{align}
where
\begin{align*}
\rho^{\mathcal{AB}_n}_{qt}(\tau) &\in  \{ \rho^{\mathcal{FG}_n}_{qt}(\tau),~ \rho^{\mathcal{G}_n}_{qt}(\tau), ~\rho^{\mathcal{OU}_n}_{qt}(\tau),~ \rho^{\mathcal{PL}_n}_{qt}(\tau) \},&
\mathcal{M}_{\mathcal{AB}_n} &\in \{\mathcal{M}_{\mathcal{FG}_n},~ \mathcal{M}_{\mathcal{G}_n},~ \mathcal{M}_{\mathcal{OU}_n},~ \mathcal{M}_{\mathcal{PL}_n} \},&
\end{align*}
with
\begin{align*}
\mathcal{M}_{\mathcal{FG}_n}= & e^{-\frac{\tau ^{2 H+2}}{H+1}} ,& \mathcal{M}_{\mathcal{G}_n}=&\exp \left[-\frac{2 \left((g \tau ) \text{erf}(g \tau ) +\frac{e^{-g^2 \tau ^2}-1}{\sqrt{\pi }}\right)}{g}\right],\\
\mathcal{M}_{\mathcal{OU}_n}=&e^{-\frac{2 \left(g \tau +e^{-g \tau }-1\right)}{g}},& \mathcal{M}_{\mathcal{PL}_n}=&\exp \left[-\frac{2 \left((g \tau +1)^{2-\alpha }+(\alpha -2) g \tau -1\right)}{(\alpha -2) g}\right].\\
\end{align*}

	\begin{figure}[ht]
		\begin{center}
			\includegraphics[width=0.40\textwidth, height=160px]{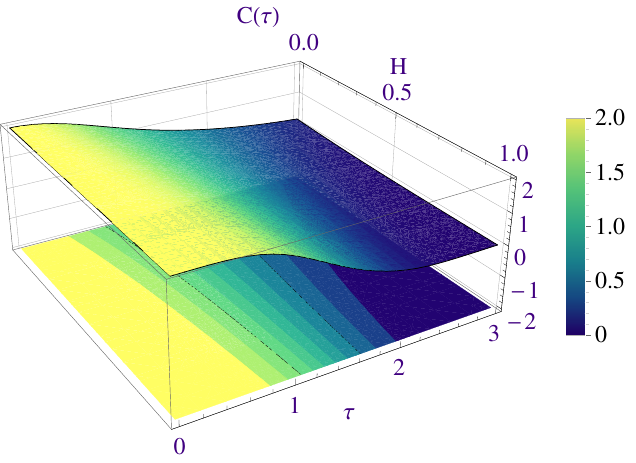}
			\put(-210,160){($ a $)} \  
			\includegraphics[width=0.40\textwidth, height=160px]{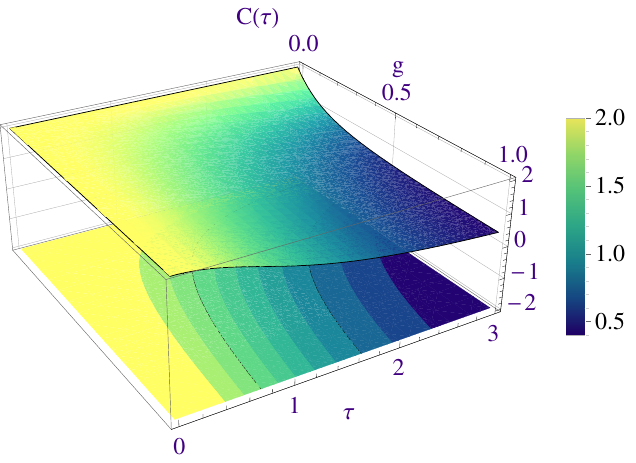}
			\put(-210,160){($ b $)}\\
					\includegraphics[width=0.40\textwidth, height=160px]{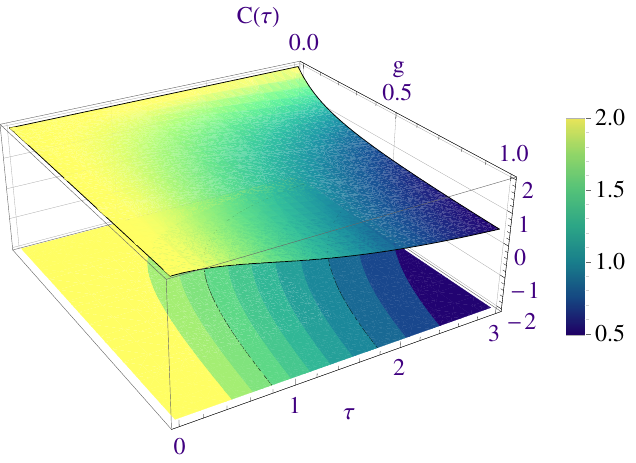}
			\put(-210,160){($ c $)} \  
			\includegraphics[width=0.40\textwidth, height=160px]{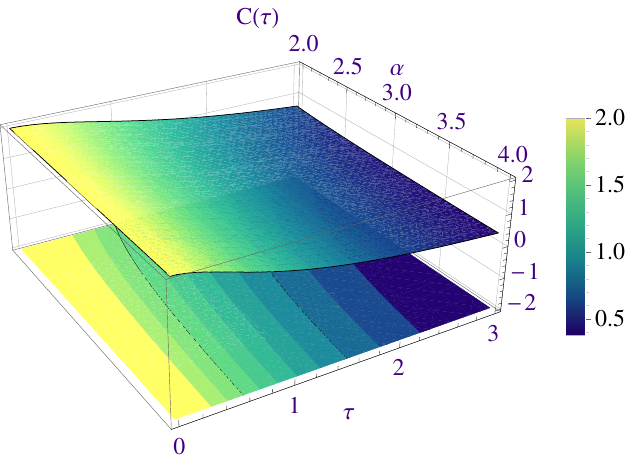}
			\put(-210,160){($ d $)}

		\end{center}
\caption{Time evolution of coherence in single qutrit system prepared in the time evolved state $\rho_{qt}(\tau)$ given in Eq.\eqref{Matrix with applied noises} when subjected to the classical field generating (a) fractional Gaussian noise when $0 \leq H \leq1$ and (b) Gaussian noise when $0 \leq g \leq1$ (c)  Ornstein Uhlenbeck noise when $0 \leq g \leq1$ and (d) power-law noise when $2 \leq \alpha \leq 4$ with $g=1$ against evolution parameter $\tau=3$.}\label{noise applied figure}
\end{figure}

Under the effects of Gaussian noises, the diagonal and off-diagonal components of the aforementioned matrix differ from those in Eq.\eqref{final density matrix value}, but they do not vanish. Therefore, the current three-level state remained coherent, even under the presence of noise. This means the single three-level system is a better resource in terms of information preservation than bipartite and tripartite quantum systems, which suffer from more loss \cite{AT1, AT2, AT3, AT4, AT5, Mazzola, Kenfack-CN, Yu, y2, y3, y4}.

\par
Using Eq.\eqref{L1 norm coherence} for the final density matrix given in Eq.\eqref{Matrix with applied noises}, the result obtained for {$\ell_1$-norm of coherence} has the following form:
\begin{equation}
C(\tau)=\frac{1}{6} \left(\left| 3+\mathcal{M}_{\mathcal{AB}_n}\right| +8\right).
\end{equation}

In Fig.\ref{noise applied figure}, the dynamics of the coherence in the final density matrix of the single qutrit system under local Gaussian noises originating from the stochastic field is depicted. The current results reflect the time evolution of the coherence when the noise phases is superimposed on the system phase. By comparing Figs.\ref{Fluctuations without noise} and \ref{noise applied figure}, one may easily determine the prevailing deteriorating character of classical noises for coherence and the revival character of the local fields. Because of the Gaussian noisy classical field, the fluctuations are eradicated after the first death. This means that the coherence revivals seen in noise-free classical fields were completely dampened in the dynamical map of the three-level system under Gaussian noises. {Therefore,} demonstrates that information transmission between the qutrit and its environment is not supported, and that information loss is irreversible. When subjected to the classical fields regulated by the Gaussian noises, the interconversion of the free and resource three-level system is completely restricted. This implies that quantum information processing involving local Gaussian noisy fields will be a delicate operation with a high risk of failure. As shown in \cite{ Mazzola, Kenfack-CN}, this characteristic has also been proved in quantum correlations and the survival of coherence in classical fields with non-Gaussian noises, causing total or partial non-local correlation losses. Each noise, as well as its associated parameters, has a different ability to suppress coherence throughout time. In the specified Hurst exponent ($H$) range, the decay is substantially higher and faster. When compared to $\mathcal{FG}_n$, the relative loss in the cases of $\mathcal{G}_n$, $\mathcal{OU}_n$, and $\mathcal{PL}_n$ is significantly reduced for both high and low values of the noisy parameters. It is worth noting that the decay behavior of $\mathcal{FG}_n$ differs significantly from that of other existing noises and, there is no evidence of a significant rise in decay when the parameter $H$ is increased. This is in direct contrast to the latter included noises when the slopes change towards higher decay as the noisy parameters are increased. The decay caused by the $\mathcal{PL}_n$ seems similar to that of $\mathcal{FG}_n$, however, the intrinsic role of their corresponding parameters $\alpha$ and $H$ seems the opposite. Besides, the $\mathcal{FG}_n$ is followed by the $\mathcal{PL}_n$ in producing greater dephasing effects. As in the cases of, $\mathcal{G}_n$ and $\mathcal{OU}_n$ noises, when $g$ approaches zero, the initial coherence remains preserved for indefinite intervals. The saturation values for all the included noises remained constant, implying a comparable relevant Gaussian character. Based on the existing findings, it is possible to predict that the non-local correlations and information decay caused by these noises will be monotonic rather than having revivals. 
\subsubsection{Classical field with $\mathcal{FG}_n$}
The dynamics of the single qutrit system originating from the classical field under $\mathcal{FG}_n$ is briefly investigated here. The impact of current noise is applied by averaging the final density matrix in Eq.\eqref{final density matrix under Gaussian process} over the noisy phase with the $\beta$-function from Eq.\eqref{Beta function of FN}. In Eq.\eqref{Matrix with applied noises}, we find that the diagonal and off-diagonal terms are non-vanishing. This means the qutrit system is still coherent under $\mathcal{FG}_n$. The density matrix states calculated for many quantum systems studied in \cite{Rossi}, contradict this, where many elements of the final density matrix of the systems vanished, resulting in larger or complete decoherence. The following analytical results are obtained using Eq.\eqref{purity} and Eq.\eqref{von Neumann entrophy} for the system's final density matrix:
\begin{figure}[ht]
		\begin{center}
			\includegraphics[width=0.40\textwidth, height=160px]{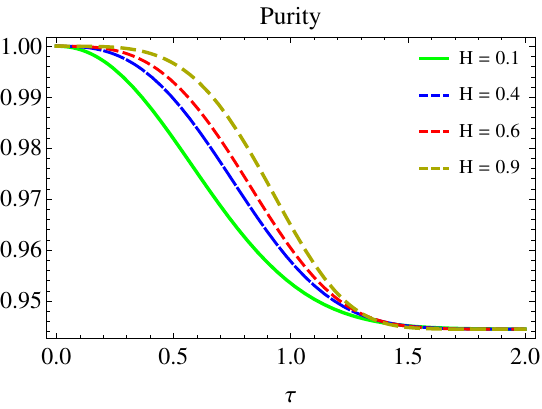}
			\put(-200,160){($ a $)} \quad \quad  
			\includegraphics[width=0.40\textwidth, height=160px]{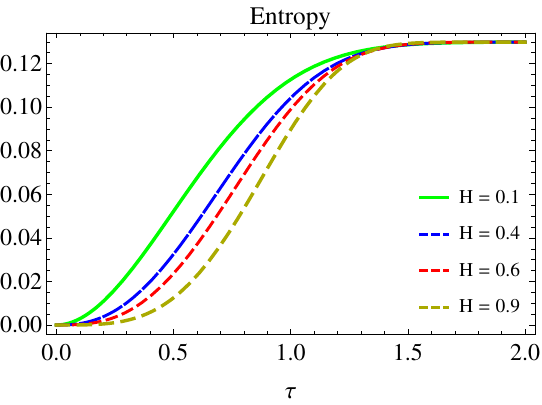}
			\put(-200,160){($ b $)}

		\end{center}
		\caption{Time evolution of (a) purity and (b) von Neumann entropy as functions of $H$ versus $\tau$ in a single qutrit system when subjected to the classical field generating fractional Gaussian noise.}\label{FGNgraphs}
\end{figure}
\begin{align}
\mathcal{P}_r(\tau)=&\frac{1}{18} \left(17+\cosh[\frac{2(\tau ^2)^{1+H} \cos[2 (1+H) \text{Arg}[\tau ]]}{1+H}]-\sin[\frac{2 (\tau ^2)^{1+H} \cos[2 (1+H) \text{Arg}[\tau ]]}{1+H}]\right),
\end{align}
\begin{align}
\begin{split}
\mathcal{V}_e(\tau)=&\frac{1}{6} \nu_1\left(-3 \nu_2+\sqrt{\nu_2+8 \nu_3}) \log[\frac{1}{6}(3-\nu_1 \sqrt{\nu_2+8 \nu_3}]\right)
\\-&\frac{1}{6} \nu_1\left(3 \nu_2+\sqrt{\nu_2+8 \nu_3})\log[\frac{1}{6}(3+\nu_1 \sqrt{\nu_2+8 \nu_3}]\right),
\end{split}
\end{align}
where
\begin{align*}
\nu_1=&e^{-\frac{2 \tau ^{2+2 H}}{1+H}}, & \nu_2=&e^{\frac{2 \tau ^{2+2 H}}{1+H}},\\
\nu_3=&e^{\frac{4 \tau ^{2+2 H}}{1+H}}.&
\end{align*}
Fig.\ref{FGNgraphs} explores the dynamics of the purity and von Neumann entropy for a single qutrit system under the local $\mathcal{FG}_n$. By comparing Figs.\ref{Fluctuations without noise}, \ref{noise applied figure} and \ref{FGNgraphs}, the destructive nature of the $\mathcal{FG}_n$ towards the revivals and preservation of the coherence and information is enough obvious. Purity and coherence, because of $\mathcal{FG}_n$, attain final saturation values after undergoing maximum decay. It is important to recall that present saturation levels only reflect a small partial loss of coherence and information. This is both startling and counterintuitive to most previous results for systems with multiple qubits or qutrits \cite{Rossi, Kenfack-CN, BenedettiC}, where a greater loss of information has been observed. This means that once the information is lost, then this noise does not facilitate the repeated interchange of information between the system and the environment. One might deduce that the information degradation is irreversible and cannot be reversed, as also shown in \cite{Rossi, Kenfack-OU, ATTA-GN, ATTA-PLFG, ATTA-PROBE}, where a temporary reversible decay occurred. The dynamics of the bipartite and tripartite states under $\mathcal{OU}_n$, pure and mixed Gaussian noises have the same monotonic qualitative decay, however, with different decay levels or reaching complete separability \cite{Rossi, Kenfack-OU, ATTA-GN, ATTA-PLFG, ATTA-PROBE}. We find it distinct that, for increasing choices of $H$, the slopes shift from green to red end. This implies the supporting nature of the parameter $H$, for memory properties of the environments and the opposite results have been obtained while discussing different noise parameters such as in \cite{Kenfack-OU, BenedettiC, Rossi, ATTA-GN} where decay increases with the increase in the noisy parameters. All parameter values overlap at the peaks and minima, indicating that both measures suggest a single saturation level{, therefore, predicting good agreement between them} The current noise parameters do not affect the loss in this case, but the noise phase has an enormous impact. Adjusting parameters has only a minor impact on the preservation duration, and any choice of $H$ does not guarantee that the coherence and information will be preserved. Because of the discrete nature, we found no ultimate solution or ideal parameter fixing to avoid the $\mathcal{FG}_n$ noisy detrimental effects. Avoiding classical environments with discrete Brownian motion of the relative constituents is the only method to reduce this type of noise. The $\beta$-function for $H=\{0.1, 0.5, 0.9\}$ is computed to be $\beta=\{2.08854,2.66667,3.66548\}$. Within classical environments with $\mathcal{FG}_n$ disorders, these precise values will be useful in optimizing quantum correlations and coherence survival time.

\subsubsection{Classical field with $\mathcal{G}_n$}
Performing the average of the final density matrix given in Eq.\eqref{final density matrix under Gaussian process} over the noise phase with $\beta$-function given in Eq.\eqref{Beta function of GN} results in the dynamics of the single qutrit system under $\mathcal{G}_n$. From the final density matrix given in Eq.\eqref{Matrix with applied noises}, it is readily deducible that the system remains coherent under the current noisy configuration because of the non-vanishing terms.
\begin{figure}[ht]
		\begin{center}
			\includegraphics[width=0.40\textwidth, height=160px]{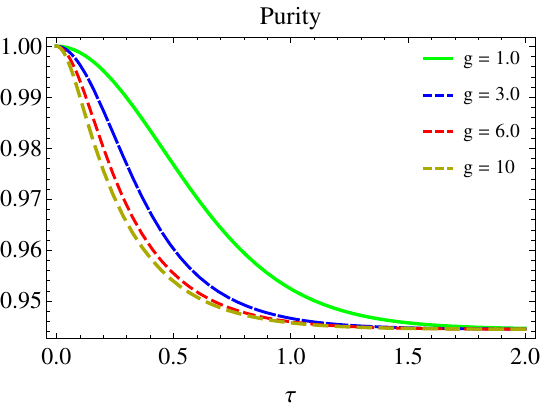}
			\put(-200,160){($ a $)} \quad \quad   
			\includegraphics[width=0.40\textwidth, height=160px]{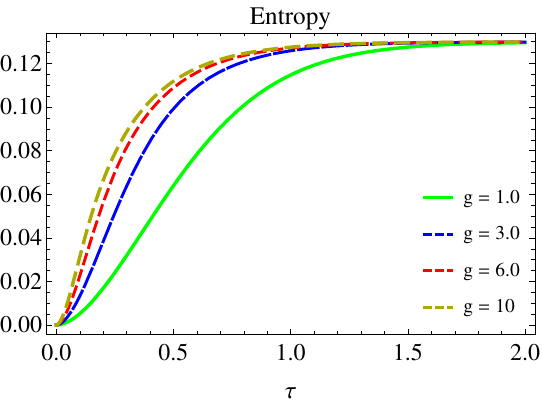}
			\put(-200,160){($ b $)}

		\end{center}
\caption{Time evolution of (a) purity and (b) von Neumann entropy as functions of $H$ versus $\tau$ in a single qutrit system when subjected to the classical field generating Gaussian noise.}\label{GN noise figure}
\end{figure}
In contrast, bipartite and tripartite states given in \cite{Rossi, AT1, AT2, AT3, AT4, AT5} remained more fragile to the local noisy environments and become less coherent shortly as compared to the current three-level system. From Eq.\eqref{purity} and \eqref{von Neumann entrophy}, the analytical results for the current configuration are:
\begin{align}
\mathcal{P}_r(\tau)=&\frac{1}{18} \left(17+e^{\frac{\frac{4 \left(1-e^{-g^2 \tau ^2}\right)}{\sqrt{\pi }}-4 g \tau  \text{Erf}[g \tau ]}{g}}\right),
\end{align}
\begin{align}
\mathcal{V}_e(\tau)=&\eta_1 \left(\left(-3 \eta_2+\sqrt{\eta_3}\right) \log\left[\frac{1}{6} \left(3-\eta_1\sqrt{\eta_3}\right)\right]-\left(3 \eta_4+\sqrt{\eta_3}\right) \log\left[\frac{1}{6} \left(3+\eta_2\sqrt{\eta_3}\right)\right]\right),
\end{align}
where
\begin{align*}
\eta_1=&\frac{1}{6} e^{-\frac{4 e^{-g^2 \tau ^2}}{g \sqrt{\pi }}-4 \tau  \text{Erf}[g \tau ]}, & \eta_2=&e^{\frac{4 e^{-g^2 \tau ^2}}{g \sqrt{\pi }}+4 \tau  \text{Erf}[g \tau ]},\\
 \eta_3=&8 e^{\frac{8 e^{-g^2 \tau ^2}}{g \sqrt{\pi }}+8 \tau  \text{Erf}[g \tau ]}+e^{\frac{4 \left(\frac{1+e^{-g^2 \tau ^2}}{\sqrt{\pi }}+g \tau  \text{Erf}[g \tau ]\right)}{g}}, & \eta_4=&e^{-\frac{4 e^{-g^2 \tau ^2}}{g \sqrt{\pi }}-4 \tau  \text{Erf}[g \tau ]}.
\end{align*} 
Fig.\ref{GN noise figure} shows the dynamics of purity and von Neumann entropy for a single qutrit system when subjected to the classical field with $\mathcal{G}_n$. The system's time evolution is investigated further for various $g$ values {against} $\tau$. By comparing Figs.\ref{Fluctuations without noise}, \ref{noise applied figure} and \ref{GN noise figure}, one can deduce the dominating deteriorating character of the $\mathcal{G}_n$ to lower preservation capacity and vanishing revival feature of the environments.
Due to $\mathcal{G}_n$, the initially encoded purity and coherence in the system showed monotonic decline rather than showing any rebirths. As a result, the classical random field with $\mathcal{G}_n$ does not allow information from the environment to flow back into the system. One can deduce that the degradation produced by $\mathcal{G}_n$ is irreversible. The observed qualitative behavior for bipartite and tripartite states is consistent with previous results obtained under various kinds of Gaussian noises studied in \cite{Rossi, Kenfack-OU, ATTA-GN, ATTA-PLFG, ATTA-PROBE}. However, the related quantitative analysis has many differences, such as greater loss and preservation time. When $g$ increases, the slopes move from the green to the red end. For large values of $g$, this means a higher occurrence of purity and coherence loss. Due to $\mathcal{FG}_n$, the decay encountered is incompatible with this behavior. It is important to note that under $\mathcal{G}_n$, the coherence and information are not lost entirely and reach a final saturation level after maximal decay. This property of the single qutrit system, which shows minimal partial loss rather than complete, is a useful resource that contradicts most previous findings obtained for bipartite and tripartite qubit systems in \cite{Rossi, Kenfack-OU, Yu, Mazzola, ATTA-GN, ATTA-PLFG, ATTA-PROBE}, where the quantum systems become easily decoherent. According to both measures, the saturation levels for each value of the noise parameter meet at the same height, implying that the decay levels are the same. The measures' maxima and minima are comparable, showing that the results are consistent. In addition, the $\mathcal{G}_n$ noise phase is less decoherent towards the coherence and information decay than the $\mathcal{FG}_n$ noise phase. Unlike the $\mathcal{FG}_n$, the $\mathcal{G}_n$ exhibits flexible noise parameter range values. This makes it easier to characterize the classical environments with $\mathcal{G}_n$ for optimal longer quantum correlations and coherence preservation time. For $g=\{1, 3,10\}$, the corresponding $\beta$-functions for the current noise have the values $\beta=\{1.43679, 1.81194, 1.94358\}$.

\subsubsection{Classical field with $\mathcal{OU}_n$}
This section involves the noisy effects due to $\mathcal{OU}_n$ by averaging the final density matrix given in Eq.\eqref{final density matrix under Gaussian process} over the noise phase having the $\beta$-function from Eq.\eqref{Beta function of OU}. The final density matrix obtained for the three-level system under $\mathcal{OU}_n$ agrees with those {obtained} under the $\mathcal{FG}_n$ and $\mathcal{G}_n$ and shows that the system remains coherent.
\begin{figure}[H]
		\begin{center}
			\includegraphics[width=0.40\textwidth, height=160px]{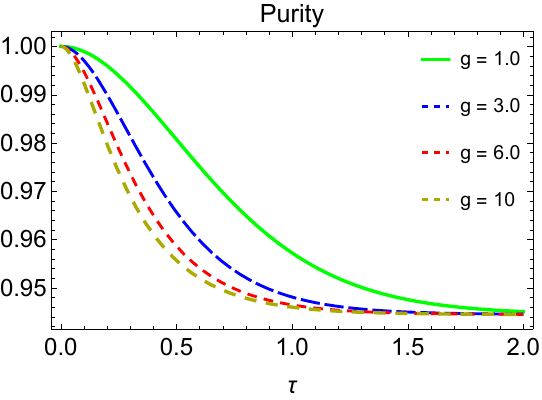}
			\put(-200,160){($ a $)} \quad \quad  
			\includegraphics[width=0.40\textwidth, height=160px]{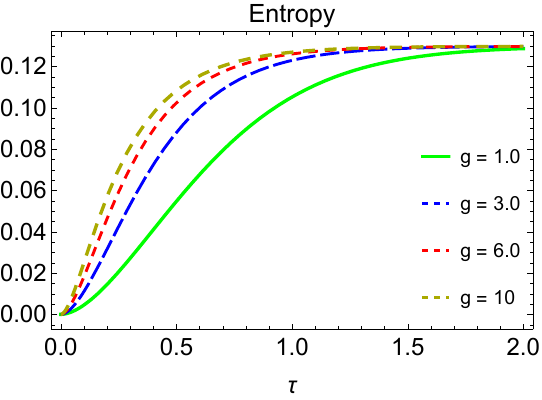}
			\put(-200,160){($ b $)}

		\end{center}
\caption{Time evolution of (a) purity and (b) von Neumann entropy as functions of $H$ versus $\tau$ in a single qutrit system when subjected to the classical field generating Ornstein Uhlenbeck noise.}\label{OU NOISE graphs}
\end{figure}
The analytical results for the $\mathcal{P}_r(\tau)$ and $\mathcal{V}_e(\tau)$ are obtained from Eq.\eqref{purity} and \eqref{von Neumann entrophy} and are followed as:
\begin{align}
\mathcal{P}_r(\tau)=&\frac{1}{18} \left(17+e^{\frac{4-4 e^{-g \tau }-4 g \tau }{g}}\right),
\end{align}
\begin{align}
\mathcal{V}_e(\tau)=&-\frac{3-\gamma_1}{6} \log\left[\frac{3-\gamma_1}{6}\right]-\frac{3+\gamma_1}{6} \log\left[\frac{3+\gamma_1}{6}\right],
\end{align}
where
\begin{align*}
\gamma_1=&e^{-\frac{4 e^{-g \tau }}{g}-4 \tau } \sqrt{e^{\frac{4}{g}+\frac{4 e^{-g \tau }}{g}+4 \tau }+8 e^{\frac{8 e^{-g \tau }}{g}+8 \tau }}, &
\end{align*}
Fig.\ref{OU NOISE graphs} shows the dynamics of the purity and von Neumann entropy for the single qutrit system when coupled to the classical field generating $\mathcal{OU}_n$. The degrading quality of the $\mathcal{OU}_n$ is shown by comparing the initial purity and coherence with the latter. The loss caused by the current noise has resulted in monotonous functions over time with no revivals. 
The current monotonic decay under $\mathcal{OU}_n$ contradicts the findings of bipartite and hybrid qubit-qutrit state dynamics given in \cite{Rossi, BenedettiC} where evident revivals of coherence have been detected. The initial encoded purity, coherence, and information are not fully lost, and the saturation threshold is reached. The measures' maximum and minimum values are comparable, and there is a single saturation level for all $g$ values. By comparing Figs.\ref{Fluctuations without noise}, \ref{noise applied figure} and \ref{OU NOISE graphs}, it is easy to determine that this noise has a dominant character to suppress oscillation and preservation capacity of the system. We noticed that as compared to the previously investigated systems in \cite{Kenfack-OU, Buscemi, Rossi-2, ATTA-PLFG, ATTA-PROBE}, the single qutrit system exhibited superior preservation capacity. Aside from that, raising the noise parameter $g$ caused the initial encoded purity, coherence, and information to decay faster. As seen, the slopes move towards the red end with increasing values of $g$, suggesting greater degradation. However, by limiting $g$ as minimal as possible, the optimal smaller decay can be produced. In contrast to $\mathcal{FG}_n$, the present noise phase has shown to have a lower deteriorating character for the memory properties of the system, as the preservation time encountered in the current case is longer. Like the $\mathcal{G}_n$, the $\mathcal{OU}_n$ has the same large range in terms of $g$. The dephasing effects because of the superposition of the $\mathcal{OU}_n$ over the system's phase is lesser than that of the $\mathcal{FG} n$. This is owing to $\mathcal{OU}_n$'s exploitable noise phase, which was not possible in the case of $\mathcal{FG}_n$. With $g=\{1,3,10\}$, the corresponding $\beta$-function amounts as $\beta=\{1.13534, 1.66749,1.9\}$.
\subsubsection{Classical field with $\mathcal{PL}_n$}
To evaluate the degrading effects of the $\mathcal{PL}_n$ over the dynamics of the single qutrit state, we perform an average of the final density matrix given in Eq.\eqref{final density matrix under Gaussian process} over the noise phase with $\beta$-function given in Eq.\eqref{Beta function of PLn}. The structure of the final density matrix under the current noise ensures that the three-level system does not become entirely decoherent. By using the Eq.\eqref{purity} and \eqref{von Neumann entrophy}, the corresponding analytical results for the $\mathcal{P}_r(\tau)$ and $\mathcal{V}_e(\tau)$ are followed as:
\begin{align}
\mathcal{P}_r(\tau)=&\frac{1}{18}\left (17+e^{4\delta_1-4 \tau -\frac{4 ((1+g \tau )^2)^{1-\frac{\alpha }{2}} \cos[(-2+\alpha ) \text{Arg}[1+g \tau ]]}{g (-2+\alpha )}}\right),
\end{align}
\begin{align}
\begin{split}
\mathcal{V}_e(\tau)=&-\frac{1}{6} \left(3-e^{-4 \delta_2} \sqrt{8 e^{8 \delta_2}+e^{4 (\delta_2+\delta_1)}}\right) \log\left[\frac{1}{6} \left(3-e^{-4 \delta_2} \sqrt{8 e^{8 \delta_2}+e^{4 (\delta_2+\delta_1)}}\right)\right]\\&-\frac{1}{6} \left(3+e^{-4 \delta_2} \sqrt{8 e^{8\delta_2}+e^{4 (\delta_2+\xi_1)}}\right) \log\left[\frac{1}{6} \left(3+e^{-4 \delta_2} \sqrt{8 e^{8\delta_2}+e^{4 (\delta_2+\delta_1)}}\right)\right],
\end{split}
\end{align}
where
\begin{align*}
\delta_1=&\frac{1}{g (-2+\alpha )},& \delta_2=&\tau +\frac{(1+g \tau )^{2-\alpha }}{g (-2+\alpha )}.
\end{align*}
\begin{figure}[ht]
		\begin{center}
			\includegraphics[width=0.40\textwidth, height=160px]{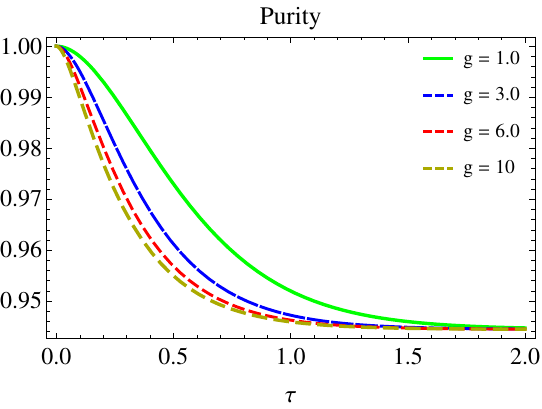}
			\put(-200,160){($ a $)}\quad \quad    
			\includegraphics[width=0.40\textwidth, height=160px]{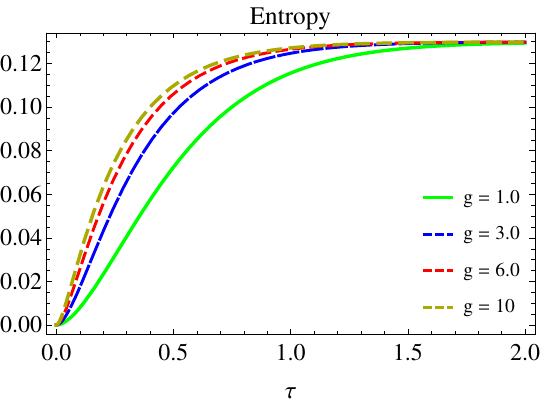}
			\put(-200,160){($ b $)}\\
						\includegraphics[width=0.40\textwidth, height=160px]{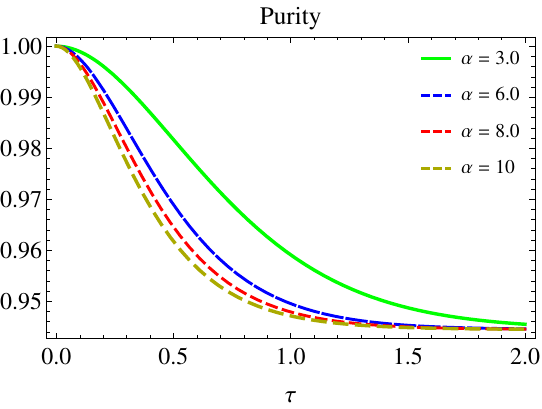}
			\put(-200,160){($ c $)} \quad \quad  
			\includegraphics[width=0.40\textwidth, height=160px]{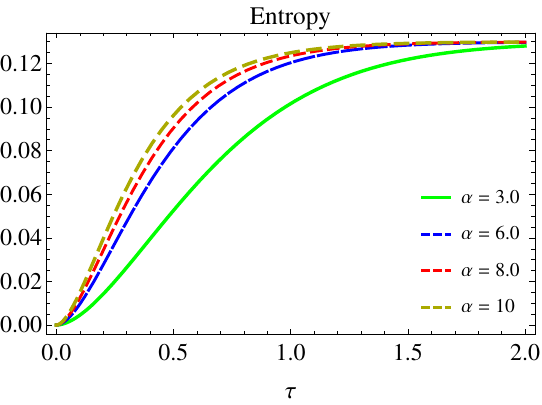}
			\put(-200,160){($ d $)}

		\end{center}
		\caption{Upper Panel: Time evolution of (a) purity and (b) von Neumann entropy as functions of $g$ versus $\tau$ in a single qutrit system when subjected to the classical field generating power-law noise when $\alpha=3$. Bottom panel:  Time evolution of {(c) purity and (d) von Neumann entropy}  as functions of $\alpha$ versus $\tau$ in a single qutrit system when subjected to the classical field generating power-law noise when $g=0.5$.}
\label{PL noise graphs}
\end{figure}

Fig.\ref{PL noise graphs} shows the time evolution of the purity and von Neumann entropy for the single qutrit system when subjected to a classical random field with $\mathcal{PL}_n$. In the current case, the dynamics of the system is investigated under two different noisy parameters, namely $g$ (upper panel) and $\alpha$ (bottom panel). By comparing Figs.\ref{Fluctuations without noise}, \ref{noise applied figure} and \ref{PL noise graphs}, the dissipative capability of the $\mathcal{PL}_n$ in terms of two noisy parameters to disappear revivals and lower the initial encoded coherence and information can be validated. Large values of $g$ have degraded purity, coherence, and information more than the parameter $\alpha$. For $g$, the slopes for purity and von Neumann entropy reach saturation values faster than for $\alpha$. Aside from the decaying nature of the $\mathcal{PL}_n$, the smaller partial loss rather than complete decay cannot be overlooked. This directly opposes most of the prior findings for various quantum systems where, in most cases, complete separability reaches, as discussed in \cite{Kenfack-OU, Rossi, Rossi-2, Weinstein, Hao, ATTA-GN, ATTA-PLFG, ATTA-PROBE}. All slopes for various noise parameter values reach a single saturation level, although at different intervals. As a result, there appears to be a strong link between the measures for demonstrating consistency and agreement in the results. As the values of both parameters increase, the slopes move from green to red end, implying that the decay rate increases. {In the current cases}, this qualitative behavior is comparable to that of the $\mathcal{OU}_n$ and mixed Gaussian noise; nevertheless, the decay levels encountered for the single qutrit system differ significantly from those previously investigated \cite{Rossi, Kenfack-OU, ATTA-PLFG}. There was no evidence of revivals in the dynamical map of the system, and both noisy parameters showed a monotonic {decline in coherence}. As a result, there is no way for information from the environment to flow back into the system, contradicting the findings in \cite{BenedettiC, Shamirzaie, Arthur} where strong backflow of the information into the system has been observed. Because of this noise, purity, coherence, and information are permanently lost rather than experiencing periodic transitory deterioration. In the case of $\mathcal{PL}_n$, the $\beta$-function is characterized by two noisy parameters, $g$ and $\alpha$. With $\alpha=3$, the $\beta$-function ranges as $\beta=\{1.333, 1.71429, 1.90476\}$ for $g=\{1,3,10\}$. By keeping $g=0.5$ and $\alpha=\{3,5,10\}$, the relative $\beta$-function has values as $\beta=\{1.0, 1.41667,1.75098\}$.
\subsection{Relative dynamics}
The present section explains the comparative dynamical map of the coherence and information encoded in the system under the present Gaussian noises. Here, we intend to provide the adjustment of the noise parameters to reduce the dephasing effects and increase the coherence span time. For this purpose, we assumed the noise parameters at the lowest and highest values of the corresponding range.
\begin{figure}[ht]
		\begin{center}
			\includegraphics[width=0.40\textwidth, height=160px]{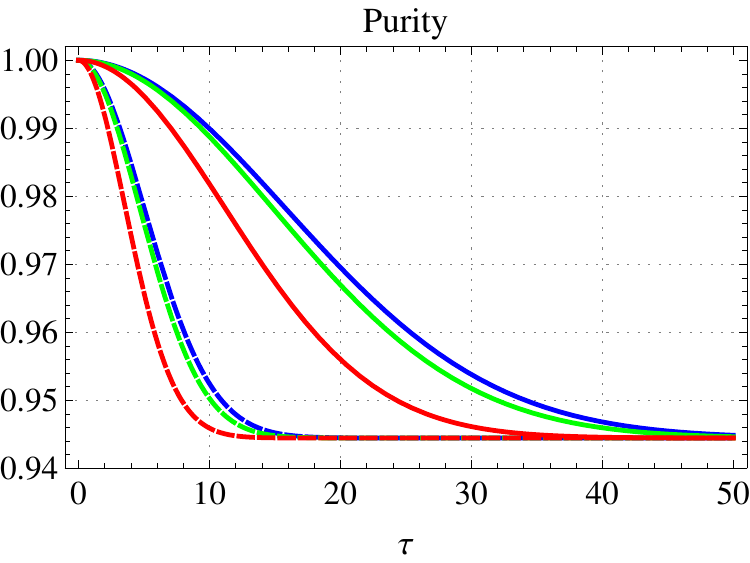}
			\put(-200,160){($ a $)} \quad \quad  
			\includegraphics[width=0.40\textwidth, height=160px]{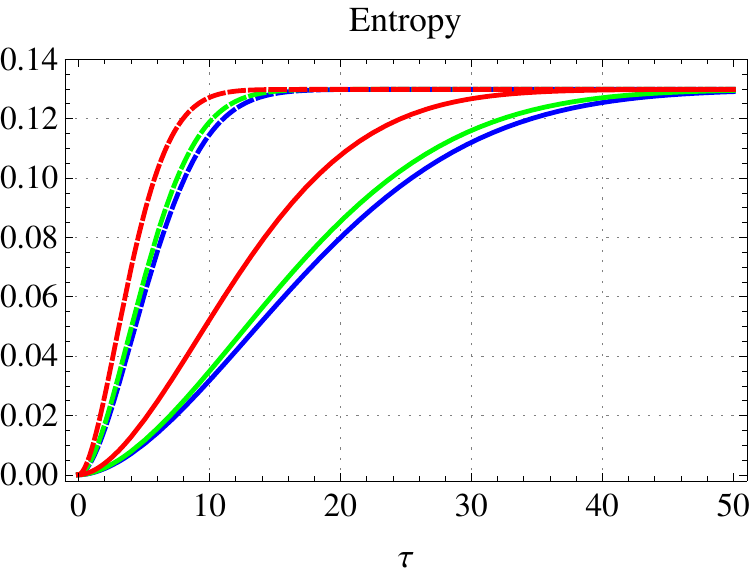}
			\put(-200,160){($ b $)}
			\end{center}

\caption{Prolong preservation of (a) purity and (b) von Neumann entropy  as functions of $g$ versus $\tau$ in a single qutrit system under Gaussian (green), Ornstein Uhlenbeck (blue) and power-law noise (red) stemming from the classical field when $g=10^{-3}$ (non-dashed) and $g=10^{-2}$ (dashed).}\label{Joint noises}
\end{figure}

Fig.\ref{Joint noises} evaluates the time evolution of the purity (a) and von Neumann entropy ($b$) for the single qutrit system under the presence of $\mathcal{G}_n$ in green, $\mathcal{OU}_n$ in blue, and $\mathcal{PL}_n$ in red slopes. We mainly focus on protecting purity, coherence, and information for a large duration. Following this, we have set the noise parameter $g=10^{-3}$ in non-dashed and $10^{-2}$ in dashed slopes {when} $\tau=50$. Note that the $\mathcal{FG}_n$ is excluded from the current study due to its discrete nature (where $0<H<1$). We discovered that the current quantitative behavior of purity and von Neumann entropy differs significantly from that observed when $g$ is large. The preservation duration of the phenomenon is substantially longer for minor values of $g$, as shown. The qualitative degradation behavior is monotone, as it was in the prior situations. In comparison, $\mathcal{OU}_n$ followed by $\mathcal{G}_n$ has had less degrading effects on the purity, coherence, and information survival over a long period. Finally, the $\mathcal{PL}_n$ is found to be the most harmful to the dynamics of the purity, coherence, and information, with saturation levels reaching earlier, especially for large $g$ values. The single qutrit system's quantitative degradation is minimal and partial and in contrast, complete coherence losses are observed in different quantum systems under different Markovian and non-Markovian noises, for example, those given in \cite{Rossi, Kenfack-OU, Arthur, Yu, ATTA-PLFG, ATTA-PROBE}. Most significantly, we found the decay rate is greatly regulated by altering the values of $g$, which directly increases as this parameter is increased. Regardless of the preservation duration and parameter values, all noises can induce a similar amount of decay. This strongly suggests the relevance of the Gaussian nature of the noises. As shown, following maximum decay, the slopes under all the noises remained at the same elevation level.

\section{Conclusion}\label{section4}
When a single qutrit system is exposed to a classical fluctuating field, the symmetrical dynamics of having preserved purity and coherence are studied. The classical fields are driven by a pure Gaussian process, generating several types of Gaussian noise. Besides, we also distinguished between noisy and noiseless local fields. Finally, the degree of pureness, coherence, and information preserved by the single qutrit system are determined using purity and von Neumann entropy.
\par
 We show that except for the $\mathcal{FG}_n$, all the noises have displayed similar degrading behavior in terms of purity, coherence, and information. The decay rate for the current Gaussian noises increased as the noise parameters were raised. On the other hand, as $H$ rises, {purity and coherence} become robust initially in the case of $\mathcal{FG}_n$. This qualitative behavior of $H$ is unlike that of any other noise parameter previously investigated. In each case, the saturation levels were consistent. This suggests the relevance of the noise phases having the same Gaussian nature and causing an equal amount of decay. Most significantly, the preservation time remained greater with $\mathcal{OU}_n$ for small $g$ and under $\mathcal{PL}_n$ for small $\alpha$. In the case of measures, purity and von Neumann entropy were found in good agreement. Over an equivalent time duration, the maxima and minima of both metrics are perfectly concordant. Therefore, purity and von Neumann entropy are accurate indicators of the initial purity, coherence, and information encoded in a quantum system. Finally, under current Gaussian noises, {purity and coherence} suffer a lesser loss, which is affected significantly by noise parameter values. In contrast to the smaller partial loss of coherence in the current three-level system, non-Gaussian noises appear to destroy coherence over time in quantum systems with a higher number of qubits and qutrits. To reduce this deterioration, the Gaussian noisy parameters should be kept as low as possible. In particular, for $g=10^{-3}$ and $g=10^{-2}$ the coherence, information, and in turn, coherence can be preserved for enough long interaction time.

\section{Acknowledgments}
S. M. Zangi is extremely grateful for help and support of Prof. Bo Zheng.
\section{Data availability statement}

The data that support the findings of this study are available upon reasonable request from the

authors.

\end{document}